**RESEARCH ARTICLE**      **OPEN ACCESS**

# RESIDENTIAL SMART PLUG WITH BLUETOOTH COMMUNICATION

**Thales Ruano Barros de Souza[1], Gabriel Góes Rodrigues[2], Luan da Silva Serrão[3], Renata do Nascimento Mota Macambira[4] and Celso Barbosa Carvalho[5]**

[1, 2, 4, 5] Federal University of Amazonas – UFAM. Manaus-Amazonas, Brazil.
[3] Northern University Center – UNINORTE. Manaus-Amazonas, Brazil.

Email: thalesrbs@ufam.edu.br, rodrigues.goes.g@gmail.com, luan.serrao@hotmail.com, reatamota@gmail.com, celsocarvalho75@gmail.com

**ABSTRACT**

Electricity forms the backbone of the modern world but increasing energy demand with the growth of urban areas in recent decades has overwhelmed the current power grid ecosystem. So, there is a need to move towards a more efficient and interconnected smart grid infrastructure. The growing popularity of the Internet of Things(IoT) has increased the demand for smart and connected devices. In this work we developed a hardware device based on the ATmega2560 microcontroller that can estimate the power consumption and control the state of electro-electronic devices interconnected to it through Bluetooth wireless technology. The developed hardware is a smart plug focusing on smart home applications. As a result, by using a smartphone device with Bluetooth communication, one can control and measure electrical parameters of the interconnected electro-electronic hardware such as the RMS (Root Mean Square) current and RMS power been consumed. The obtained results showed the technical viability in the construction of energy consumption measuring device using modules and components available in the Brazilian market.

---

## I. INTRODUCTION

Users of the electricity supply networks are largely unaware of the electricity consumption of their household electrical appliances, and the lack of technical knowledge about their home's electrical installations and appliances is frustrating.

This paper seeks to use the concept of IoT to help solve this problem by implementing a device capable of estimating the energy consumption of a home appliance to which it is connected and transmitting this information via Bluetooth wireless communication to a smartphone device. Bluetooth has been used as it is currently one of the most widely used technologies in solutions that integrate wireless sensors and mobile devices.

IoT is a paradigm that has the idea of integrating objects into people's daily lives, using wired and wireless sensors, tracking technologies and actuator networks, bringing practicality to users' lives [1-5]. Within this view, the concept of IoT can allow access to a huge amount of data generated daily by such objects and, therefore, enables the creation of differentiated services by public agencies and private companies [6].

In this work we intend to use the concept of IoT integration in the construction of energy consumption meter device that using Bluetooth wireless communication is able to monitor the consumption of an electro-electronic device to which it is connected. In the work was used shelf hardware, available in the national market, to build a prototype of such device, showing technical feasibility.

## II. THEORETICAL FOUNDATION

This section presents the basic concepts related to Bluetooth technology and the current sensor employed in the development of this research.

## II.1 BLUETOOTH

The first steps in the development of Bluetooth technology were taken by Ericsson in 1994 in search of a low cost and low power solution between mobile communication devices and their accessories. The potentiality of the project caught the attention of companies in the industry that in 1998 joined together forming the Bluetooth Special Interest Group (Bluetooth SIG), a group of 5 companies responsible for developing Bluetooth standards, licensing technology and trademark to other companies. device manufacturers using the technology. The following year, the first version of the Bluetooth specification was released, demarcating connection patterns, devices, procedures and security [7].

Bluetooth technology works on an ad hoc wireless network, meaning you don't need a pre-existing infrastructure made up of routers or access points. This Bluetooth network without infrastructure is called piconet. A piconet must have a device identified as master, responsible for controlling, synchronizing transmissions and registering new Bluetooth network devices. A maximum of 7 devices called slaves can also actively participate in the network, which communicate directly with the master device [7].

The master device can also register other devices to be added to the network in the future as needed, paired with them. Devices that are not associated with any piconet are identified as standby.

Figure 1 shows an HC-05 hardware module used in this article. The HC-05 provides Bluetooth wireless communication and also has a serial interface that integrates with external devices such as computers or microcontrollers intuitively, cheaply and affordably.

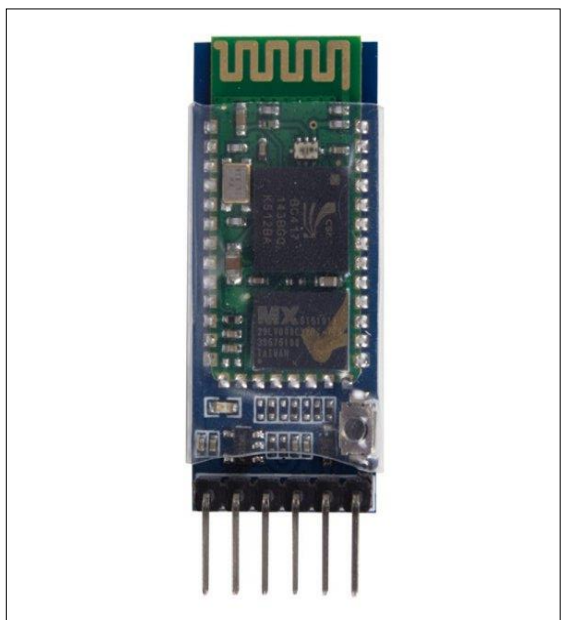

Figure 1: HC-05 Module.
Source: Authors, (2019).

The HC-05 has two operating modes, corresponding to Bluetooth operating modes, namely master mode and slave mode. An HC-05 module configured in master mode is capable of creating a piconet. To configure the module and change its parameters, it is necessary to use AT mode, provided that the module is open to receive, via serial communication, control commands.

## II.2 ACS712 SENSOR

The ACS712 sensor is a device capable of reading AC (Alternative Current) or DC (Discrete Current) current, enabling integration with other industrial, commercial or communication electronic devices. The ACS712 is composed of a hall effect sensor connected to a copper track, which once a current travels, excites the sensor, generating an analog signal of voltage proportional to the read current [8].

This device is available in the market in 3 versions that differ in both the current supported by each model and the accuracy of each, with models of 5, 20 and 30 A (amps). The current supported by the ACS712 is inversely proportional to the sensitivity of its reading.

The ACS712 comes in an 8-pin Small Outline Integrated Circuit (SOIC) package, with pinouts implemented as shown in Figure 2 [8]:

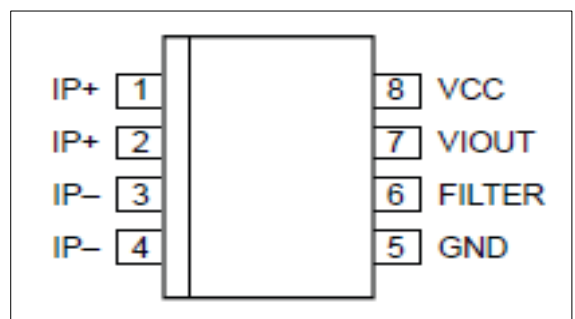

Figure 2: IC ACS712 Pinout Diagram.
Source: [8].

An important feature of this device is described in its technical specification, the IC (Circuit Integrated) operates with output that depends on the applied supply voltage. Causing reading to be severely impaired if any variation in input voltage occurs.

The ACS712 (Figure 3) is sold in IC form and can be integrated with hardware prototype microcontrollers such as the one proposed in this article.

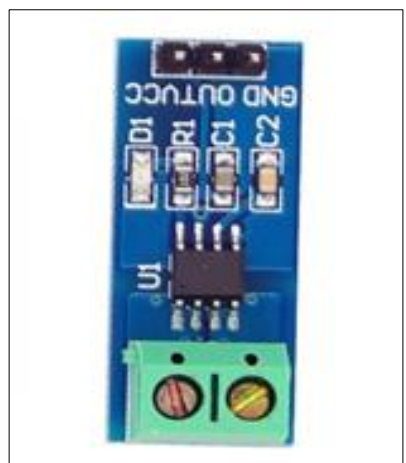

Figure 3: ACS712 in module.
Source: Authors, (2019).

The ACS712 operates so that if supplied with a 5V voltage source and if a current flowing from pins 3 and 4 to pins 1 and 2 of the IC (Figure 4) of an intensity equal to the rated current is applied at Its input is expected to read 0.5 V at the output. If the current has the reverse direction (which flows from pins 1 and 2 to pins 3 and 4 of the IC). A 4.5V reading is expected as shown in the Figure 4:

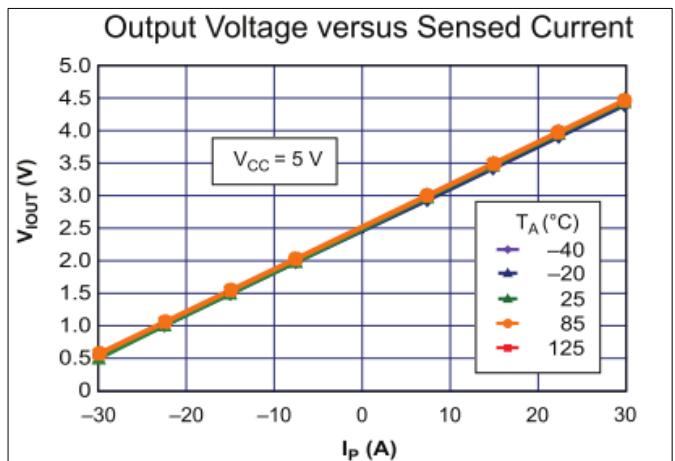

Figure 4: Description of output voltage as a function of current applied to ACS712 sensor.
Source: [8].

In the image above it is observed that it is possible to obtain a satisfactory level of stability for the system at the output of the ACS712 despite temperature variations, which is desirable because it allows the proposed prototype to be robust in various climates and installation conditions.

### III. LITERATURE REVISION

Researchers have been using wireless technologies such as Zigbee, Wi-Fi, and Bluetooth for smart socket design. Some techniques have also been suggested for reducing power consumption by employing smart outlets.

The paper [9] introduces the technique of identifying devices in real time using smart sockets. The proposed system uses an Arduino microcontroller and an ESP8266 module, respectively for Wi-Fi processing and communication. The authors use a classification technique to identify the device using the consumed power data.

The proposal of [10] identifies the electrical events using the noise or the pulse transition during the equipment's electrical state switching. Noise during switching (on / off) of an application is detected, classified and associated with an electrical event or equipment using supervised learning techniques.

In the work implemented by [11] a new architecture is proposed to reduce energy use in domestic environments. The paper proposes the use of multiple Energy Management Devices (EMDs) to form the basic building block of the architecture. The blocks are connected by an EMD hub outside the network.

For [12] proposed a smart socket solution where a Power Line Monitor (PLM) is employed on each wall socket. PLMN data is reported every 10 seconds.

The nPlug developed by [13] uses analytical intelligence to determine peak demand periods, load unbalance, and it is possible to schedule high power electrical operation for off-peak hours without manual user intervention.

A power management system developed by [14] uses Bluetooth to detect user presence and control devices in user space using smart outlets. A smartphone app and a web service show the energy consumed. The article also proposed an energy saving algorithm to control the power consumed by the device plugged into the smart outlet. The system was developed and tested for a period of three months and the average energy savings were 31.3% and 15.3% for computer and lighting respectively.

In [15] an intelligent plug was proposed where an algorithm controls the voltage applied to the load instead of turning it off. This reduces the instantaneous power consumed on passive loads resulting in a low peak demand. The system uses the Zigbee communication protocol and a voltage control circuit registering an 18% reduction in peak demand. The system, however, has some limitations such as the considerable increase in THD (Total Harmonic Distortion) during voltage control.

In [16] the authors designed and developed a smart plug using Bluetooth for power consumption monitoring using a mobile app. The paper also compared some of the current smart sockets with the proposed device.

For [17] presented a smart plug solution that can monitor as well as control devices using a web application. This system uses Wi-Fi as a wireless communication protocol and a relay to control the state of devices. Measurement accuracy was computed and an error of less than 0.5% was reported by the authors.

For [18] Developed control hardware that analyzes through a sensor network the levels of illuminance in a given environment, later controlling the power applied to the lighting circuit, helping to reduce the consumption of electricity.

### IV. METHODOLOGY

In this project we want to implement a system capable of reading the current, voltage and power consumed by a load connected to it. This system should be able to estimate these variables for any type of load (resistive, reactive or switched) with the least possible interference with the read values. In order to achieve these goals, the block diagram illustrated in Figure 5 describing the main components of the system was developed:

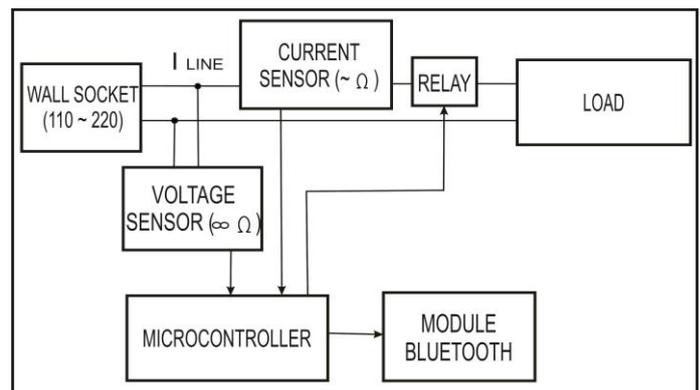

Figure 5: Block diagram of the hardware solution.
Source: Authors, (2017).

The block diagram shows the following components:

• Wall Socket - Represents the power input to which the system will be connected and which will power both the load and the active electronic monitoring and control device components. Since the proposed device is intended for use in a home environment, voltage in the range 110 to 220V RMS is assumed for calculation and circuit design purposes;

• Current Sensor - Represents a device capable of converting the value of the AC current that feeds the load into a voltage signal compatible with the chosen microcontroller. For this project the ACS712ELECTR-30A-T sensor produced by AllegroTM was chosen. For prototyping purposes, a development module provided by the company was used to encapsulate the chip.

• Voltage sensor - Sensor capable of converting a high voltage level that feeds the load linearly into a low voltage level compatible with the chosen microcontroller. This device was designed and implemented in a module with passive components in a process that will be described later in the article.

• Microcontroller - ATMELA2560 was chosen from ATMEL Corporation which works in conjunction with Arduino Bootloader and integrated into a development board to facilitate the prototyping process.

• Relay - Represents the switching device used to cut the load supply. In the article the relay is controlled directly by the microcontroller, without the use of drivers. For this article we used the SRD-05VDC-SL-C, compatible with 5V operating voltage and 10 A maximum load current. The relay development board was used for simpler prototyping.

• Bluetooth Module - The Bluetooth module is a component capable of communicating with the microcontroller using serial communication interface, as well as transmitting information obtained from the microcontroller to other devices such as smartphones. The module used in the article was HC-05.

• Load - Due to the fact that the prototype of this article is intended for domestic use, it is expected that the loads found by the system will be mostly composed of switched sources, except for a few cases, because due to computerization, the most commonly used domestic loads are of this nature such as mobile phone chargers, computer supplies, LED lamps, etc.

### IV.1 CURRENT READING WITH ACS712 MODULE

Using the code shown in Figure 6 it is possible to observe the output of the ACS712 module in real time through the serial communication present on the microcontroller development board. In addition, the IDE provided by the Arduino platform enables real-time visualization of the collected signal through its Plotter Serial tool. This allows the code and the reading module to be validated in their operation prior to the definitive implementation of the prototype.

```
1  //Rotina simples de leitura de porta analógica
2  void setup() {
3    // Inicialização serial a um baudrate de 250000
4    Serial.begin(250000);
5  }
6
7  void loop() {
8    // realiza a leitura da entrada analógica
9    int sensorValue = analogRead(A0);
10   // escrita do valor de sensor lido na
11   // interface serial
12   Serial.println(sensorValue);
13 }
```

Figure 6: Simple analog read code.
Source: Authors, (2019).

For a current reading through the ACS712 module it is necessary that the module has its reading terminals, pins 1-2 and 3-4 connected in series with the load for which the reading process of the module is to be performed. chain. In Figure 7, these pins will not be visible as the modular version of the ACS712 appears, where C.I. pins 1-2 and 3-4 are equivalent to the two (2) connection terminals located on the right side of the module.

The circuit allows a current through the ACS 712 sensor module to be converted to an analog voltage signal at output pin 2 (pin OUT) with respect to pin reference 3, GND. This voltage can be converted to a digital signal via the A/D converter (Analog / Digital) integrated in the microcontroller used. The A/D converter can be accessed via the "AIN0" pin on the Arduino Mega development board, which can be configured via C language programming code.

In order to validate the current reading method, the current module was connected in series with a fully resistive load electric device (15W soldering iron) in order to verify the current consumed by it non-invasively. The resistive element chosen was used as the load on the circuit shown in Figure 7.

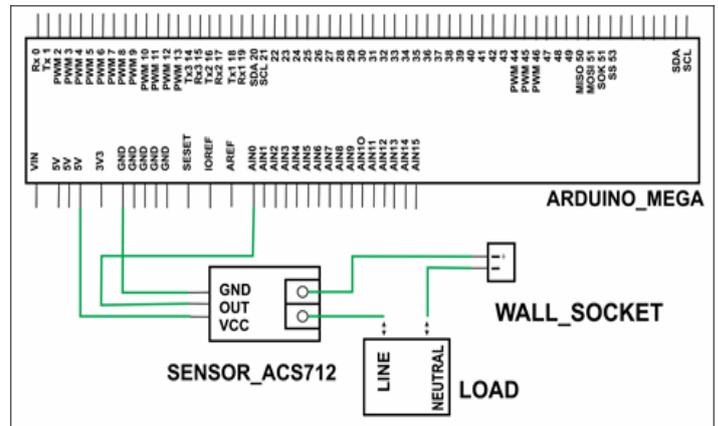

Figure 7: Current Reading Circuit.
Source: Authors, (2019).

Using the graphical construction tools present in the Arduino platform, we obtained the signal illustrated in Figure 8. This is the signal of the ACS712 module output voltage, which when interpreted by the Arduino A / D converter assumed values between 510 and 519 within an integer range from 0 to 1024 corresponding to the voltages of 0V and 5V respectively. Therefore, the above values must be converted to voltage values through the relationship shown in Equation (1):

Conversion of the measured voltage

a) $$V = \left(\frac{D}{1024}\right) * 5000 mV \qquad (1)$$

Where D is the integer value obtained by the A/D converter and V is the estimated voltage at the ACS712 sensor output. With this information and knowing that the relationship between V and the current that flows through the terminals of the ACS712 is directly proportional, and specified by the manufacturer, it is possible to estimate the electric current consumed by the load. Since the ACS712 module used in this article is of model ACS712ELCTR-30A-T, it is observed that the relationship between output voltage and read current is 66mV / A, according to the documentation provided by the manufacturer. Equation 2 illustrates how current can be calculated, where S is the sensitivity of the module as specified by the manufacturer.

Measured current conversion

b) $$I = \frac{\left(\frac{D}{1024}\right)*5000mV}{S} \qquad (2)$$

Figure 8 illustrates the current behavior read by the ACS712 module. It is observed that the current oscillates around the value 39A which is not the value actually consumed by the load. This is due to an offset present in the sensor.

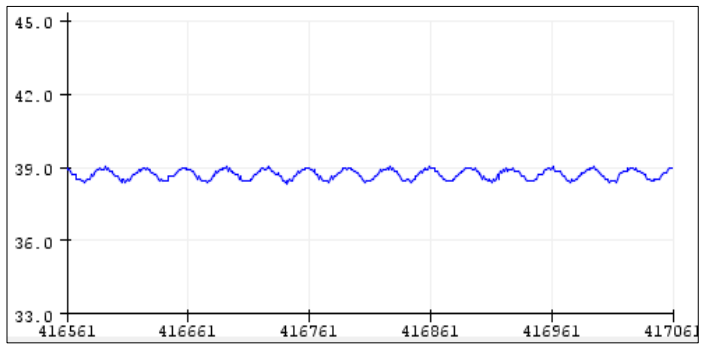

Figure 8: Offset measured current signal.
Source: Authors, (2019).

To correct the offset, a moving average algorithm using the last 70 samples was used to determine the average current value. The resulting signal is shown in Figure 9.

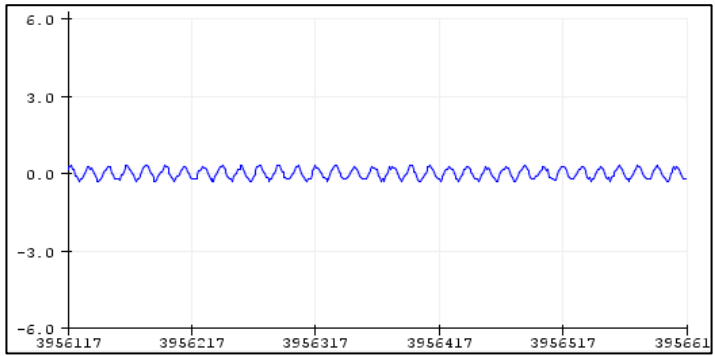

Figure 9: Current reading without offset.
Source: Authors, (2019).

### IV.2 VOLTAGE READING

Unlike current reading, where a commercial module facilitates prototyping work, no module was found that met the prototype needs and was compatible with the Arduino microcontroller. Therefore, it was necessary to develop a module capable of providing isolation between the line voltage and the electronic circuits present in the system and to transfer the voltage information read on the line in a linear, instantaneous and with low current consumption.

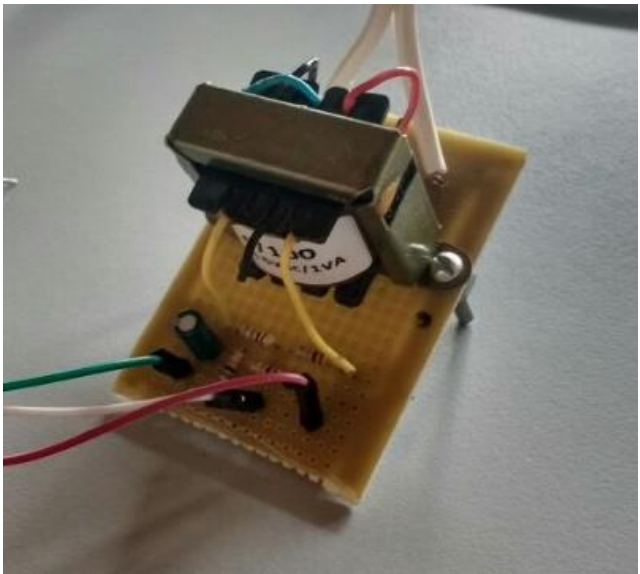

Figure 10: Voltage sensor module developed.
Source: Authors, (2019).

According to Figure 10 the designed voltage sensor module has three distinct parts:

1. The transformer that is responsible for lowering the voltage and isolating the circuit from voltage fluctuations. This has a voltage reduction factor of 100: 6.

2. In Figure 11, resistors R1 and R2 are responsible for dividing the voltage from the transformer so that it is within the operating values of the Arduino microcontroller.

3. Also in Figure 11, resistors R3 and R4 are responsible for raising the signal to a DC level equal to half of the microcontroller read voltage range, so that the generated voltage signal, including negative voltage values, visible to the ADC (Analog to Digital Converter) converter of the Arduino microcontroller. Capacitor C1 has the function of stabilizing this DC voltage level avoiding to generate sudden variations that may be available in the Arduino ADC input.

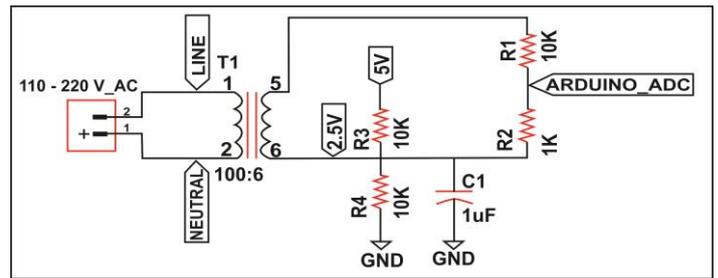

Figure 11: Voltage sensor module circuit diagram.
Source: Authors, (2019).

For the voltage sensor module to be used as a voltage reading device, a linearity constant that relates the actual voltage and the voltage read by the Arduino must be found (Equation 3). Assuming, for calculation purposes, that the peak voltage value of 220VAC (Volts Alternating Current) is present at the input of the voltage sensor module. You can calculate the read voltage value and the linearity constant that relates it to the voltage observed by the microcontroller. Assuming an ideal transformer:

Obtaining the actual voltage

c) $$V_{out} = \frac{6}{100} * V_{in} = 13{,}2V \qquad (3)$$

The voltage V_out corresponds to the peak voltage VAC observed at the transformer output. Thus, assuming the resistors R1 and R2 to be ideal gives Equation 4:

Voltage divider

d) $$V_{div} = 13{,}2V * \frac{R_2}{R_1+R_2} = 1{,}2V \qquad (4)$$

The voltage $V_{div}$ div is observed by the voltage divider at the circuit output, but it is necessary to sum the DC level in resistors R3 and R4 in order to obtain the value detected by Arduino. This is because, in the previous calculation we are taking as reference this level DC and not the level DC in relation to the ground. Therefore, the actual voltage available on Arduino is shown in Equation (5).

Real tension

e) $$V_{adc} = 1{,}2V + DC = 3{,}7V \qquad (5)$$

The linearity relationship between the line voltage and the voltage observed by Arduino was obtained according to Equation (6). However, due to the nonlinear characteristics of the components used, there was a considerable difference between the value obtained and that observed in practice. Therefore, it was decided to perform empirical tests to obtain a satisfactory linearity constant for the proper functioning of the system.

Linearity constant

f) $$S = \frac{V_{arduino}}{V_{in}} = \frac{3,7V}{220V} = 16,8 mV \quad (6)$$

From the value obtained in the above calculation, an iterative improvement search process was performed in order to find a value. The RMS voltage of a 127V electrical outlet was verified using a multimeter. Then, the same reading was performed through the microcontroller in order to refine the obtained proportionality constant. The obtained empirical proportionality constant is shown in Equation (7).

Proportionality constant

g) $$S = 5,2 mV \quad (7)$$

Similar to the procedure performed to correct the obtained current, the voltage signal obtained was treated. Using a vector of determined size to fix the number of samples, an arithmetic average of the values is calculated and this average is used to correct the stress value.

### IV.3 BLUETOOTH MODULE CONFIGURATION

To use the Bluetooth module, it was necessary to perform previous configurations of: i) slave mode operation; ii) configuration of the module name in the network; iii) access password setting. These setup operations can only be performed in AT mode.

To access AT mode the Bluetooth module is connected to the Arduino microcontroller, which during this process will operate as a serial interface between the serial terminal and the Bluetooth module. Arduino serial communication / Bluetooth module is performed through the receiver (Rx) and transmitter (Tx) pins of the HC-05 module and the microcontroller pins 10 and 11, respectively. It was also necessary to activate the KeyPin pin present in the Bluetooth module. To perform the configuration, this pin must be at logic level 1 when the module is energized.

Figure 12 illustrates the wiring diagram of the connection between the Bluetooth module and the Arduino microcontroller. Importantly, in order for the configuration to work and AT mode to be successfully accessed, the HC-05 module power supply must be connected only after the Arduino pin 9 (PWM / 9) raises the KeyPindo HC-module pin. 05 to 5 Volts, otherwise the BluetoothHC-05 module will not enter AT operation mode.

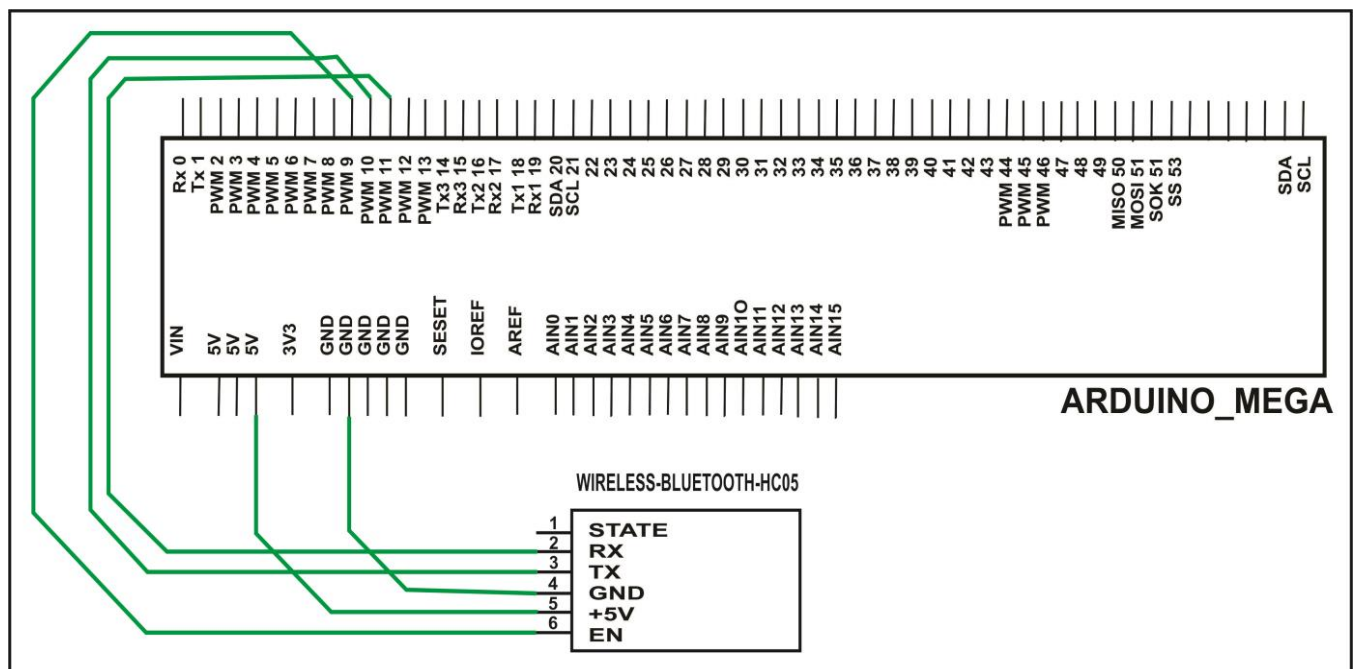

Figure 12: Electrical diagram of HC-05 module connection in AT mode.
Source: Authors, (2019).

### IV. PROTOTYPE CREATION

A prototype based on the diagram in Figure 5 was developed. The prototype has a USB port that allows communication with a computer in order to perform code loading and electrical powering of the Arduino module. The prototype also has a plug used to connect the appliance to be controlled and whose voltage and current you want to read. In addition, the prototype also has a socket that allows connection to a socket, and is compatible with voltages ranging from 110 to 220VRMS.

The prototype, whose electrical scheme is shown in Figure 13, was encapsulated in a 14.5x9.5x5.5cm plastic box, in which all components were inserted and fixed with few clearances. To the voltage sensor input was connected a plug according to the Brazilian standard (NBR14136), with 4mm diameter terminals, thus supporting up to 10A. To the output of the circuit was connected a socket of 4x2,3cm also of the Brazilian standard, supporting a maximum of 10A.

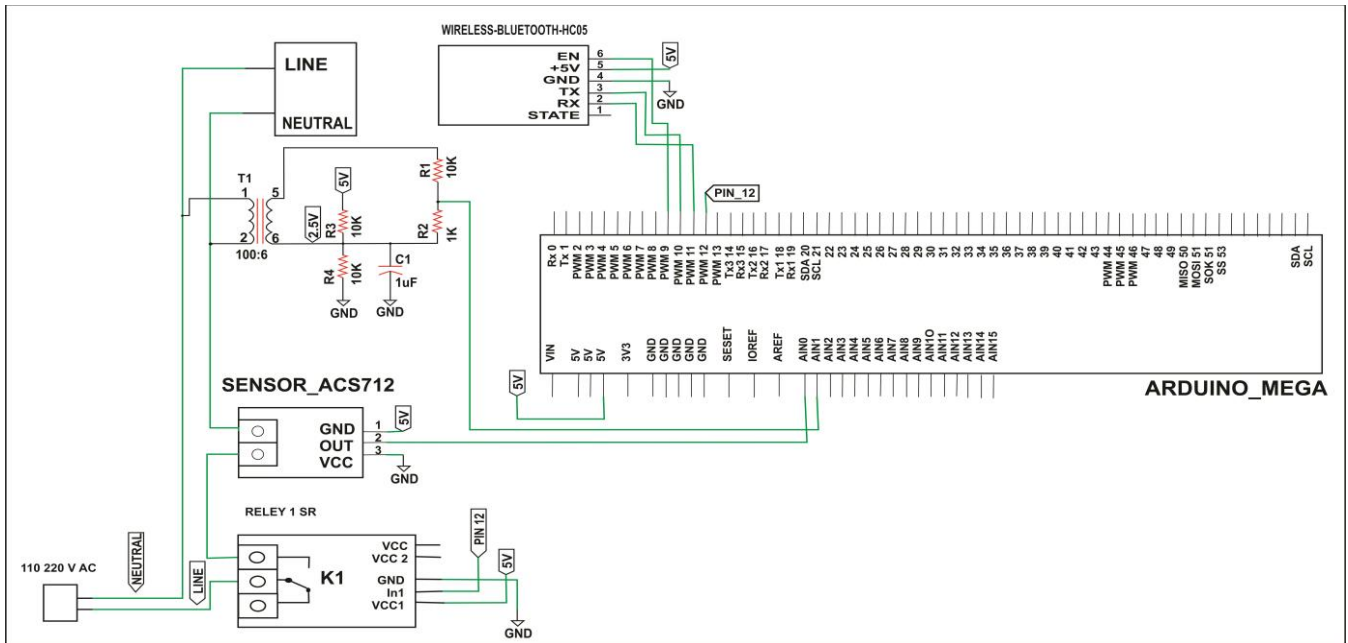

Figure 13: Electrical prototype diagram.
Source: Authors, (2019).

The components were arranged inside the box in 2 layers. In the first layer the voltage sensor, current sensor, relay and socket were arranged and in the second layer the Arduino module was positioned as shown in Figure 14:

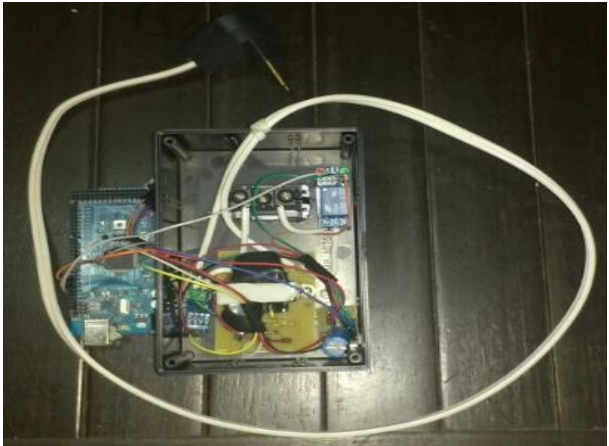

Figure 14: Open prototype.
Source: Authors, (2019).

The Bluetooth module has been fixed to the prototype side wall to minimize interference from other components and cannot be seen in the image above due to perspective.

### IV.5 INSTANT READ RELIABILITY TESTING METHODOLOGY

For a reading system as developed in this article there is a need for frequent voltage and current readings. The ideal situation is where these physical quantities are measured simultaneously, but they are measured by the Arduino microcontroller and therefore read sequentially. One possible consequence of this factor is that there is a lag between the instantaneous voltage and current reading that is not due to the circuit, but rather the small-time difference between the current/voltage measurements. To measure the influence of this effect on reading the following test protocol was performed:

1. Obtain 30 zero-pass readings of both the voltage signal and the current signal during prototype operation while feeding a resistive load and an inductive load.
2. Calculate the instant of time at which both the voltage signal and the current signal pass through zero, by determining a line representing the transition between a positive and a negative value, according to Equations 8 and 9.

Zero Crossing Voltage

h) $$Z_v = \frac{-(T_p*T_n*V_p)+(V_p*T_n^2)+(V_n*T_p^2)-(V_n*T_p*T_n)}{(V_p*T_n)-(T_n*V_n)-(T_p*V_p)+(T_p*V_n)} \quad (8)$$

Zero Crossing Current

i) $$Z_c = \frac{-(T_p*T_n*C_p)+(C_p*T_n^2)+(C_n*T_p^2)-(C_n*T_p*T_n)}{(C_p*T_n)-(T_n*C_n)-(T_p*C_p)+(T_p*C_n)} \quad (9)$$

Where:
• Positive zero crossing voltage (Vp) - last voltage value read by the microcontroller before the sine waveform presents its negative cycle;
• Negative zero traversal voltage (Vn) - first voltage value read by the microcontroller during the negative cycle of the read sine waveform;
• Positive zero-pass current (Ip) - last current value read by the microcontroller before the sine waveform presents its negative cycle;
• Negative zero-through current (In) - first value of current read by the microcontroller during the negative cycle of the read sine waveform;
• Positive zero crossing time (Tp) - time in microseconds at which the positive zero crossing voltage or current was obtained;
• Negative zero crossing time (Tn) - time in microseconds at which the negative zero crossing voltage or current was obtained.
Calculate the difference between the time points of the zero and voltage current passage, according to Equation 7.

Zero Crossing Difference

j) $$T_d = Z_c - Z_v \qquad (7)$$

1. Estimate the current and voltage lag. Knowing that for a resistive load it is expected to observe a lag of 0% and for an inductive load it is desired to observe a delay of the current in relation to the voltage, according to the following formulation:

Inductive load lag

l) $$\phi = T_d * 10^{-6} * 360 * 60 \qquad (8)$$

## IV.6 VALIDATION OF THE RMS VALUE CALCULATION METHOD

It is necessary that the method for obtaining the RMS values implemented in this article be validated and compared with other means of obtaining, for this the following procedures were performed:

1. Obtain 30 sample readings obtained during prototype operation by feeding a resistive load. The prototype uses serial communication to communicate with a personal computer. The 30 reading samples shall have information on the peak voltage beyond the time at which the reading was taken, information provided by the microsecond scale microcontroller;
2. Use of the peak voltages obtained previously to calculate the RMS voltage through the relationship shown in Equation 9, and the calculation shall be performed for 30 available voltage values;

RMS voltage

m) $$V_{rms} = \frac{V_p}{\sqrt{2}} \qquad (9)$$

3. Calculation of the mean and standard deviation of the 30 RMS values for comparison and further discussion;
4. Using a multimeter to check the available RMS voltage at the outlet where the system is powered, the reading should be taken a total of 30 times at least 1 minute intervals. The values must be recorded;
5. Calculation of the mean and standard deviation of the 30 RMS values collected from the multimeter;
6. Obtaining 30 RMS values calculated using the method implemented in the source code in Annex A provided by serial communication to a personal computer by the prototype, the values shall be recorded every 1 minute at least;
7. Calculation of the average of 30 RMS values and their standard deviation for comparison and subsequent discussion;
8. Discussion and comparison of the 3 methods.

## IV.7 RELIABILITY TEST OF ACTIVE POWER ESTIMATE

In this paper, two methods for obtaining power were used, one for obtaining active power and one for obtaining reactive power. It is useful to check if the first one is in agreement with the second, presenting identical values for a resistive load. Since these are completely different methods, this would prove that the values provided by the system actually converge at a given reading. The following test aims to verify the reliability of the RMS (reactive) and active power estimation methods:

1. Obtain 30 samples of active and RMS power readings obtained during prototype operation by feeding a resistive load in serial communication with a personal computer. Readings should be taken within a minimum of 1 minute;
2. Calculate the mean and standard deviation of the difference between readings;
3. Analyze the results.

## V. EXPERIMENTS AND RESULTS

The instant read accuracy tests followed the protocols described in Session IV.5. Values were acquired using the prototype. A total of 30 voltage and current samples together with the time the reading was obtained in microseconds from the microcontroller energization instant.

In Figure 15, besides recording the values verified in the experiment, the estimated moment in which the voltage value occurs assumes the zero value. This information is useful because when compared to the moment when the current assumes zero, one can estimate the lag between voltage and current waveforms.

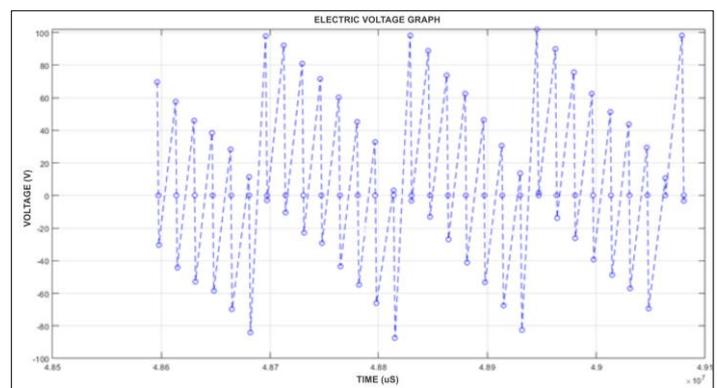

Figure 15: Voltage signal measured for resistive load.
Source: Authors, (2019).

Figure 16 represents the plot of the collected current values as well as the zero crossing time for resistive load.

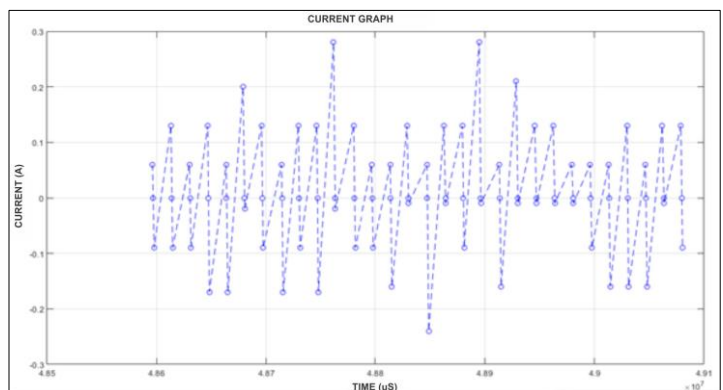

Figure 16: Current signal measured for resistive load.
Source: Authors, (2019).

Figure 17 illustrates the shifting behavior between voltage and current for the resistive load. Taking into account the mains frequency of 60 Hz, the difference between the zero crossing between the current signal and the voltage signal previously measured is calculated. Then the result is converted to microseconds to degrees.

The average of the lag values of the graph in Figure 17 is -2.59 ± 2.59 degrees, indicating a slight delay of voltage relative to current. This result does not match the behavior expected for a purely resistive load. With this, we conclude that there is an error of 0.71% in the estimation of current and voltage lag. This average has a standard deviation of 7.55 degrees.

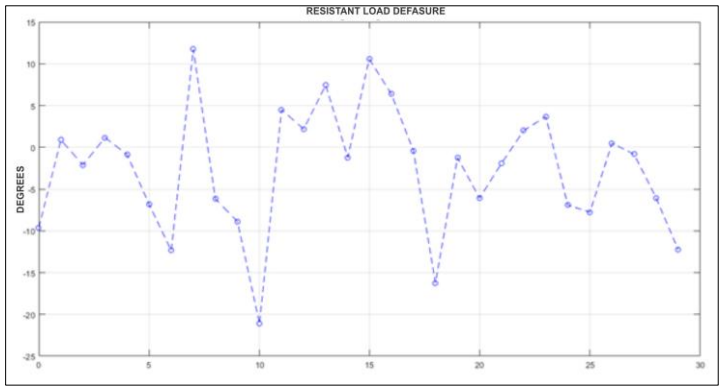

Figure 17: Resistive load lag.
Source: Authors, (2019).

Then the tests involving an inductive load were performed. For the experiment in question a 127V home fan was chosen. Due to the nature of the load, the average lag value is expected to be greater than zero, indicating a delay of current to voltage. Figure 18 shows the current signal.

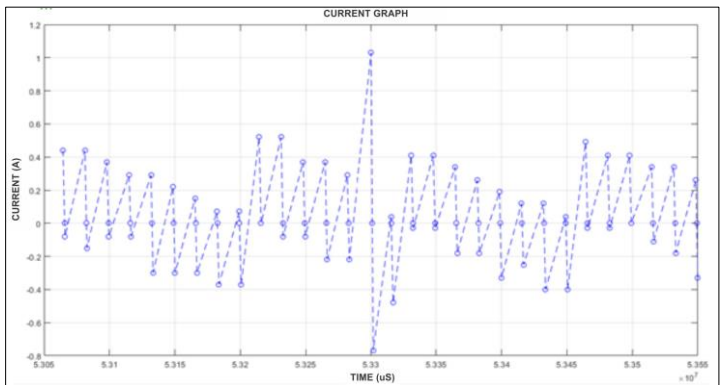

Figure 18: Current signal measured for inductive load.
Source: Authors, (2019).

Figure 19 illustrates the voltage signal behavior measured for inductive load.

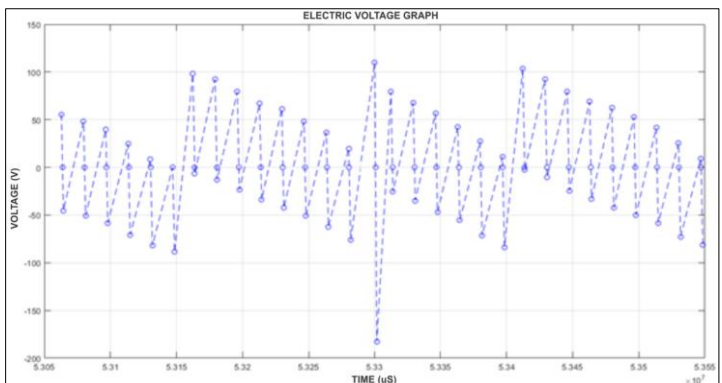

Figure 19: Voltage signal measured for inductive load.
Source: Authors, (2019).

After an analysis of the graph of Figure 20, showing the lag between voltage and current for the inductive load, it can be concluded that there was a delay of current to voltage of approximately $〚42〛^o$ on average, with a standard deviation of $7^o$, a standard deviation very close to that obtained for resistive load.

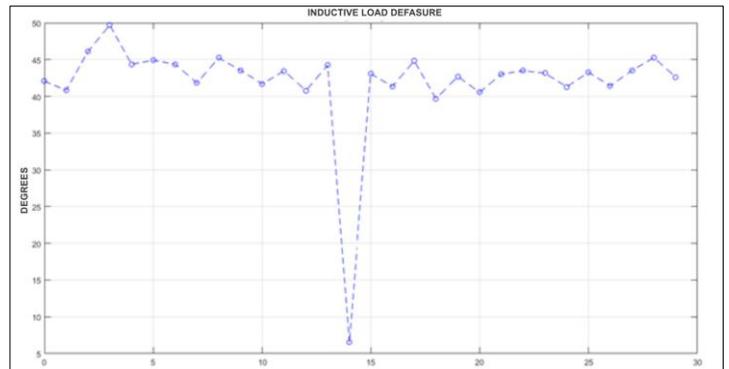

Figure 20: Inductive Load Defeasure.
Source: Authors, (2019).

From the collected data, it can be concluded that there is an error inherent in the lag detected by the system. However, this error is within a standard deviation of at most $〚7.55〛^o$, meaning that for domestic applications where accuracy in offset reading is not a critical item, these results are more than sufficient to estimate residential consumption.

The next test performed, with results presented in Figure 21, was the RMS power measurement accuracy. For this purpose, 30 samples were collected from: i) peak voltage; ii) calculated RMS voltage; (iii) Bluetooth voltage: Measured by means of the voltage sensor developed and transmitted via the Bluetooth wireless network. iv) and the RMS voltage measured by a digital multimeter.

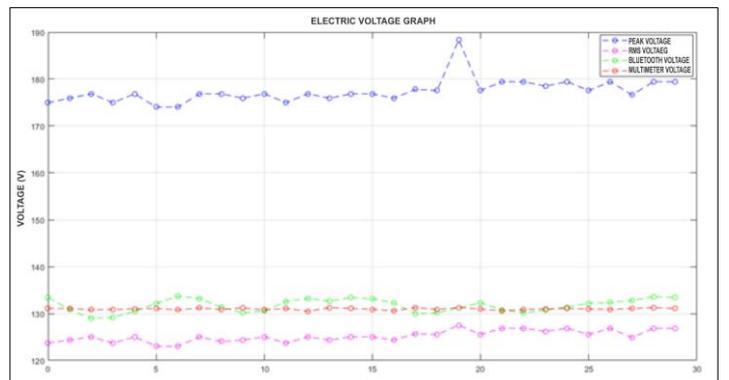

Figure 21: Stresses obtained for the calculation of RMS power.
Source: Authors, (2019).

The measured RMS voltage measured by a digital multimeter averages 131 V with a standard deviation of 0.21 V. This reading generates a base value against which the other readings are compared. The RMS voltage values calculated from peak voltages only have an average of 125.24V and a standard deviation of 1.22V. Therefore, an error of 5.76V is calculated in relation to the average obtained using the digital multimeter. The values calculated for Bluetooth voltage showed an average of 131.77V and a standard deviation of 1.41V, thus presenting an error of 0.77V in relation to the average of the values measured by the multimeter.

Thus, the technique implemented in the article for RMS value calculation actually had a higher accuracy than that obtained by the simplest method using peak voltage. This means that the prototype had an error of approximately 0.5%, which would be more than enough to effectively estimate RMS power values.

In order to verify the data obtained for active power, a resistive load connected to the system was used, since it is known that in this situation there is no delay between current and voltage and, therefore, there is no reactive component in the power making. RMS power and active power are identical. Figure 23 presents the observed data showing that RMS power values resulted in an average of 32Var and standard deviation of 0.78Var. Already the active power values showed an average of 31.56W and standard deviation of 0.77. In fact, the value of these two powers, obtained by completely different methods presented very close values. As observed in Figure 22, the difference between the two averages was only 0.56W, with a standard deviation of 0.15W. This indicates that for a resistive load the active powers and RMS are similar with an error of only 1.74%.

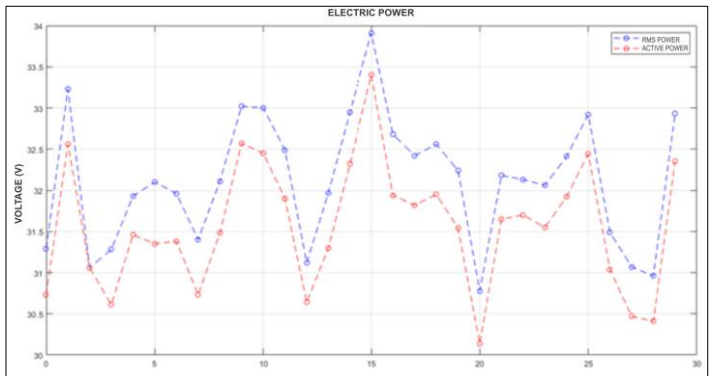

Figure 22: RMS Power and Active Power.
Source: Authors, (2019).

## VI. CONCLUSIONS

In this work a prototype of a meter of electric energy consumption of electronic devices was developed. The prototype is capable of measuring quantities such as voltage, instantaneous current and RMS of the electro-electronic device to which it is connected and transmitting via Bluetooth communication the measurements made to a Smartphone. The tests and results show that the prototype developed, both in hardware and software can estimate the values of current, voltage, apparent power, active power and reactive power satisfactorily for domestic use and with the purpose of informing the consumer about your expenses. As future work we intend to design and develop a switching power supply for the microcontroller module allowing greater energy efficiency and therefore less waste. Another possibility would be the use of instrumentation amplifiers for voltage measurement.

## VII. ACKNOWLEDGMENTS

This research, as provided for in Article 48 of Decree No. 6,008 / 2006, was funded by Samsung Eletrônica da Amazônia Ltda, pursuant to Federal Law No. 8,387 / 1991, through agreement No. 004, signed with CETELI / UFAM; This research received infrastructure support from the Higher Education Personnel Improvement Coordination (CAPES), the Amazonas State Research Support Foundation (FAPEAM) - support programs (First Projects Program (PPP)), the Technical Chamber of Reconstruction and Infrastructure Recovery (CT-INFRA) of the Ministry of Science, Technology, Innovations and Communications (MCTI) / National Council for Scientific and Technological Development (CNPq) and by the State Secretariat of Science, Technology and Innovation - Amazonas (SECTI-AM ) and Government of the Amazon State, Brazil.